\documentclass{llncs}
\usepackage{llncsdoc}

\usepackage{algorithmic, algorithm2e, color}
\usepackage{graphicx, subfigure, epstopdf}
\usepackage{amsmath, amssymb, mathtools}
\usepackage{graphicx, subfigure, epstopdf}
\usepackage{exscale,relsize}
\usepackage{amsfonts}
\usepackage{mdwlist}
\usepackage{cite}
	
\usepackage{array} 
\usepackage{url}
\usepackage{graphicx}
\usepackage{subfigure}
\usepackage{listings}
\usepackage{dsfont}
\usepackage{enumitem}
\usepackage{paralist}

\usepackage{times}
\title{Lights, Camera, Action! \\  Exploring Effects of Visual Distractions on \\  Completion of Security Tasks}
\author{Bruce Berg 
\and Tyler Kaczmarek
\and Alfred Kobsa 
\and Gene Tsudik \\} 
\institute{University of California Irvine, Irvine, CA, USA \\ \email{bgberg@uci.edu \\
tkaczmar@uci.edu\\kobsa@uci.edi\\gts@ics.uci.edu}}
\begin{document}

\maketitle

\begin{abstract}
Human errors in performing security-critical tasks are typically blamed on the
complexity of those tasks. However, such errors can also occur because of (possibly
unexpected) sensory distractions. A sensory distraction that produces negative effects
can be abused by the adversary that controls the environment. Meanwhile, a 
distraction with positive effects can be artificially introduced to improve user performance.

The goal of this work is to explore the effects of visual stimuli on the performance of security-critical tasks. To this end,
we experimented with a large number of subjects who were exposed to a
range of unexpected visual stimuli while attempting to perform Bluetooth Pairing. Our results clearly demonstrate substantially increased task completion times
and markedly lower task success rates. These negative effects are noteworthy,
especially, when contrasted with prior results on audio distractions which had
positive effects on performance of similar tasks.
Experiments were conducted in a novel (fully automated and completely unattended) experimental
environment. This yielded more uniform experiments, better scalability and
significantly lower financial and logistical burdens. We discuss this experience, including
benefits and limitations of the unattended automated experiment paradigm.
\end{abstract}

\section{Introduction}
\label{sec:intro}
It is widely believed that the human user is the weakest link in the security chain. 
Nonetheless,  human participation is unavoidable in many security protocols. Such
protocols require extensive usability testing, since users are unlikely to perform well when 
faced with overly difficult or intricate tasks. Typically, security-related usability testing 
entails evaluating human performance in a ``best-case" scenario. In other words, testing 
is usually conducted in sterile lab-like environments. 

At the same time, security protocols involving human users have become more commonplace. 
Examples include activities, such as: (1) using a personal device for verification 
of transaction amounts, (2) entering a PIN or a password and (3) solving a CAPTCHA,  
(4) comparing PINs when pairing Bluetooth devices, and (5) answering personal security 
questions. 

Since overall security of these tasks is determined by the human user (as the weakest link), 
extensive usability studies have been conducted. They aimed to assess users' ability to 
perform security tasks correctly and without undue delays, while providing an acceptable 
level of security \cite{kobsa_serial_2009} \cite{goodrich_using_2009} \cite{nithyanand_groupthink:_2010} 
\cite{kainda_usability_2009}. 

However, the focus on maximizing successful protocol completion led developers to evaluate usability 
under contrived and unrealistic settings. In practice, security tasks can take place in noisy environments. 
In real-world settings, users are often exposed to various sensory stimuli. The impact of such stimuli on 
performance and completion of security tasks has not been well studied. A particular stimulus (e.g., a fire alarm 
or flickering lights) can be unintentional or hostile, i.e., introduced by the adversary that controls the physical 
environment. Furthermore, recent emergence of Internet of Things (IoT) devices (such as smart speakers and 
light fixtures) in home and office settings creates environments where compromised (malware-infected) 
devices can expose users to a variety of visual and audio stimuli. 

There has been just one prior study that studied the effects of stimuli on the completion of security-critical tasks. 
It showed that introduction of unexpected audio stimuli during Bluetooth pairing actually improved subject performance \cite{kaczmarek_unattended_2015}. This initial result, though interesting, motivates a more thorough study 
in order to fully understand the effects of a range of unexpected (and potentially malicious) stimuli.

Since modern user-aided security protocols focus on maximizing successful outcomes
in an ideal environment, human errors are quite rare. For example, Uzun et al. \cite{Uzun_2007} assume that :
\begin{quote}
``...[A]ny non-zero fatal error rate in the sample size of 40
is unacceptable for security applications."
\end{quote} 
Consequently, numerous trials with many subjects are needed to
gather data sufficient for making claims about human error rates. The scale is further 
exacerbated by the need to test  multiple modalities, each with a distinct set of subjects. (This is because a 
given subject is less likely to make a similar mistake twice, even under different conditions.) Therefore, 
the number of required participants can quickly grow into hundreds, which presents a logistical 
challenge. To ease the burden of conducting a large-scale study, we designed and employed an 
entirely unattended and automated experimental setup, wherein subjects receive recorded
instructions from a life-sized projection of a video-recorded experimenter (``avatar"), 
instead of a live experimenter.
 
We extensively experimented with subjects attempting to pair two Bluetooth devices (one of which
was the subject's own device) in the presence of various unexpected visual stimuli. 
We tested a total of $169$ subjects in the fully unattended experiment 
setting.\footnote{
All experiments described in this paper were fully authorized by the Institutional Review Board (IRB).} 
We initially hypothesized 
that visual stimuli would have beneficial or facilitatory effects on subject task completion,
as was recently experienced with its audio counterpart \cite{kaczmarek_unattended_2015}.  
Surprisingly, we discovered a marked slowdown in task completion times across the 
board, and lower task success rates under certain stimuli.

The rest of the paper is organized as follows: The next section overviews related 
work and background material. Then, we present the design and setup of our 
experiments, followed by the presentation of our experimental results. Next, we derive 
conclusions and summarize lessons learned. The paper concludes with the discussion of 
limitations of our approach and directions for future work. Appendix 1 presents and analyzes
performance of subjects arriving in groups.
Appendix 2 contains the description of color spaces used to generate our stimuli.
Details on the unattended experiment setup are in Appendix 3.

\section{Background \& Related Work}
\label{sec:related}
This section overviews related work in  automated experiments, and human-assisted 
security methods.We also provide background information in psychology, particularly 
effects of sensory arousal on task performance, as well as effects of visual 
stimuli on arousal level and emotive state.

\subsection{Automated Experiments}
Other than recent results describing effects of audio distractions \cite{kaczmarek_unattended_2015},
we are unaware of any prior usability studies utilizing a fully automated and unattended 
physical environment.  

However, some prior work reinforces validity of virtually-attended remote experiments and 
unattended online surveys, in contrast with same efforts in a traditional lab-based setting. Ollesch 
et al.\cite{ollesch_physical_2006} collected psychometric data in: (1) a physically attended 
experimental lab setting and (2) its virtually attended remote counterpart. No significant differences 
were found. This is further reinforced by Riva et al.\cite{riva_use_2003} who 
compared data collected from (1) unattended online, and (2) attended offline, questionnaires. 
Finally, Lazem and Gracanin \cite{lazem_social_2010} replicated two classical social psychology 
experiments where both the participants and the experimenter were represented by avatars in 
Second Life\footnote{See \url{secondlife.com}}, instead of being physically co-present. 
Here too, no significant differences were observed.

\subsection{User Studies of Secure Device Pairing}
Secure device pairing (mostly, but not only, via Bluetooth) has been extensively researched by 
experts in both security and usability. While initially pairing, the two devices have
no prior knowledge of one another, i.e., there is no prior security context. Also, they can not
rely on either a Trusted Third Party (TTP) or a Public Key Infrastructure (PKI) to facilitate the protocol. 
This makes device pairing especially vulnerable to man-in-the-middle (MiTM) attacks. This prompted 
the design of numerous protocols requiring human involvement (integrity verification) over 
some out-of-band (OOB) channel, e.g., visual or audio comparison or copying/entering numbers.

For example, Short Authenticated String (SAS) protocols ask the user to compare two 
strings of about 20 bits each \cite{cryptoeprint:2005:424}. 

Uzun et al.~\cite{Uzun_2007} performed the first usability study of Bluetooth 
pairing techniques using SAS. It determined that the ``compare-and-confirm" method
-- which involves the user comparing two 4-to-6-digit decimal numbers and indicating 
a match or lack thereof -- was the most accurate and usable approach. 

Kobsa et al.~\cite{kobsa_serial_2009} compiled a comprehensive comparative 
usability study of eleven major secure device pairing methods. They measured task 
performance times, completion times, completion rates, perceived usability and 
perceived security. This led to the identification of most problematic as well as  
most effective pairing methods, for various device configurations.

Goodrich et al.~\cite{goodrich_using_2009} proposed an authentication protocol 
that used ``Mad-Lib" style SAS. Each device in this protocol creates a nonsensical 
phrase based on the protocol outcome, and the user then determine if the two phrases
match. This approach was found to be easier for non-specialist users.

Kainda et al.~\cite{kainda_usability_2009} examined usability of device pairing in a group 
setting. In this setting, up to 6 users tried to connect their devices to one another by participating 
in a SAS protocol. It was found that group effort decreased the expected rate of security and 
non-security failures. However, if a single individual was shown a SAS different from that of 
all others participants, the former often lied about the SAS in order to fit in with the group, 
demonstrating so-called ``insecurity of conformity.''

Gallego et al.~\cite{sadeghi_exploring_2013} discovered that subject's performance 
in secure device pairing could be improved if it were to be scored. In other words, 
notifying subjects about their performance score resulted in fewer errors.

\subsection{Effects of Sensory Stimulation}

Sensory stimulation has variable impact on task performance. This is due to many contributing factors, 
including the subject's current level of arousal. The Yerkes-Dodson Law stipulates an inverse 
quadratic relationship between arousal and task performance \cite{cohen_yerkesdodson_2011}. It implies
that, across all contributing stimulants, subjects who are either at a very low, or very high, level of arousal 
are not likely to perform well, and there exists an optimal level of arousal for correct task completion. 

An extension to this law is the notion that completion of less complex tasks that produce lower levels of 
initial arousal in subjects benefits from inclusion of external stimuli. At the same time, 
completion of complex tasks 
that produce a high level of initial arousal suffers from the inclusion of external stimuli. 
Hockey \cite{hockey_effect_1970} and Benignus et al.~\cite{benignus_effect_1975} classified this 
causal relationship by defining the complexity of a task as a function of the task's event rate 
(i.e., how many subtasks must be completed in a given time-frame) and the number of sources 
that originate 
these subtasks. External stimulation can serve to sharpen the focus of a subject at a low arousal level, 
improving task performance \cite{olmedo_maintenance_1977}. Conversely, it can overload subjects 
that are already at a high level of arousal, and induce errors in task completion \cite{harris_stress_1960}.

O'Malley and Poplawsky \cite{omalley_noise-induced_1971} argued that sensory noise affects 
behavioral selectivity. Specifically, while a consistent positive or negative effect on task completion 
may not occur, a consistent negative effect was observed for tasks that require subjects to react to 
signals on their periphery. Meanwhile, a consistent positive effect on task completion was observed 
for tasks that require subjects to react to signals in the center of their field of attention. This leads to the claim that sensory stimulation has the effect of narrowing the subject's area of attention.

\subsection{Unique Effects of Visual Stimuli}
In addition to being general external stimuli that serve to raise arousal level, 
visual stimuli, particularly colors, have social and emotional implications. Naz and Epps \cite{1} 
surveyed 98 college students about their emotional responses to five principal hues 
(red, blue, purple, green and yellow), five intermediate hues (yellow-red, green-yellow, blue-green, 
and red-purple) as well as three achromatic colors (white, gray, and black.) They found that principal 
hues are more likely to foster positive emotive responses. Furthermore, different colors within 
each group induce differing levels of arousal: some (red or green-yellow) 
increase arousal, while others (blue and green) are perceived as relaxing.

Moreover, visual stimuli were found to be dominating in multi-sensory contexts. 
Eimer \cite{eimer_multisensory_2004} showed that in experiments with tactile, visual, and audio stimuli, 
subjects overwhelmingly utilized visual queues to localize tactile and auditory events. 

\section{Methodology}
\label{sec:experiment}
This section describes our experimental setup, procedures and subject parameters.

\subsection{Apparatus}
\label{subsec:apparatus}
\begin{figure}
    \centering
    \begin{minipage}{0.5\textwidth}
\fbox{\centering
	\includegraphics[height=2.0in, width =0.9\textwidth]{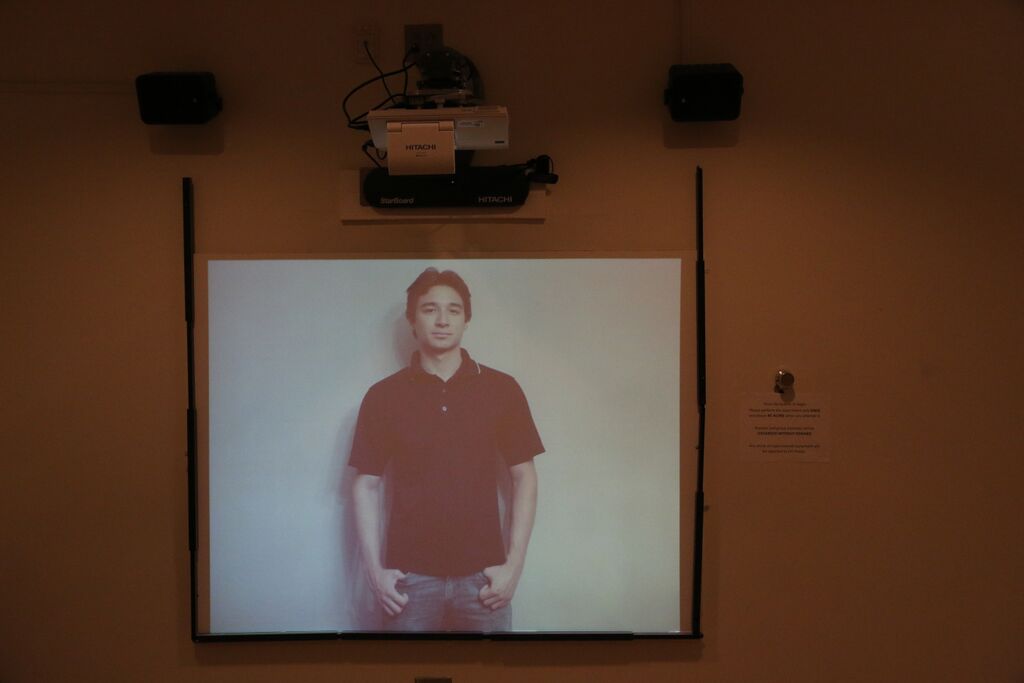}}
	\caption{\scriptsize{Experimental environment: subject's perspective}}
    \end{minipage}\hfill
    \begin{minipage}{0.5\textwidth}
\fbox{\centering
	\includegraphics[height=2.0in, width =0.9\textwidth]{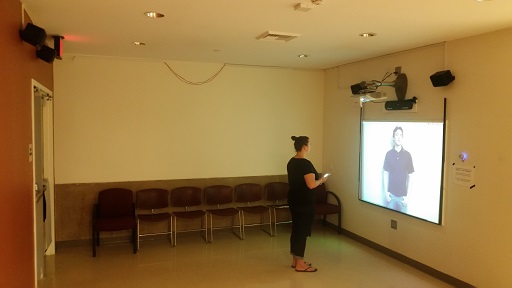}}
	\caption{\scriptsize{Experimental environment: side view}}
    \end{minipage}
\end{figure}
The experimental setting was designed to facilitate fully automated experiments with a wide range of sensory inputs. We located the experiment in a public, but low-traffic alcove at the top floor of the Computer Science Department building in a large 
public university. Figure 1 shows our setup from the subject's perspective (front view), 
and Figure 2 depicts it from the side. More photos can be found in Appendix 2.
The setup is comprised of readily available off-the-shelf components:
\begin{compactitem}
\item A 60"-by-45" touch-sensitive interactive Smartboard\footnotemark[2]
whiteboard with a Hitachi CP-A300N short-throw projector\footnotemark[2].
The Smartboard acts as both an input and a display device. It reacts to tactile input, 
i.e., the user touches its surface, similar to a large touch-screen. 
\item A Logitech C920 HD Webcam\footnotemark[2].
\item Two pairs of BIC America RtR V44-2 speakers\footnotemark[2]: 
one alongside the smartboard, and the other -- on the opposite wall. Their arrangement is 
such that the subject is typically standing in the center of the four speakers.
\item Four programmable wirelessly controllable Phillips Hue A19 LED 
lightbulbs\footnote{See: \url{meethue.com} for Hue Bulbs, \url{smarttech.com} for the Smartboard,
\url{logitech.com} for the Webcam, \url{bicamerica.com} for speakers, and \url{hitachi.com} for 
the projector.} to deliver the visual stimuli. 
\end{compactitem}
device. All prospective subjects were explicitly informed, during recruitment, that they would 
need to use their own personal device that supports Bluetooth communication. 
We could have instead provided a device to the subjects, which might have fostered a more uniform 
subject experience. However, there would have been some drawbacks: 
\begin{compactitem}
\item We wanted to avoid accidental errors due to the use of an 
unfamiliar device that might have a different user interface from that 
of the subject's own device. Mitigating this unfamiliarity would have required some training, 
which is incompatible with the unattended experiment setting. 
\item Virtually all current Bluetooth pairing scenarios involve at least one of the 
devices being owned by the person performing the pairing. Forcing subjects to use our
device would have resulted in a more contrived or synthetic experience.   
\item From a purely practical perspective, 
an unattended portable device provided by us would have been more
prone to damage or theft than other components, which are  bulky and 
attached to walls and/or ceilings.
\end{compactitem}
Not surprisingly, the majority of subjects' devices (152 out of 169) were smartphones. 
Tablets (13) and laptops (4) accounted for the rest.

Bluetooth pairing is not as common as other security-critical tasks,
such as password entry or CAPTCHA solving.
However, we believe that Bluetooth pairing is the ideal security-critical task 
for the unattended experiment setup. It is preferred to passwords and PINs 
since it does not require subjects to reveal existing, or to select new, secrets.
The security task at the core of  Bluetooth pairing involves the user comparing two 
6-digit decimal numbers -- one displayed by each device being paired -- and pressing 
a single button. This is a much more discrete and uniform activity than solving CAPTCHA-s, 
which vary widely in terms of difficulty and require higher-resolution displays as well as
more extensive user input. These factors, even without external stimuli, would yield 
large variations in error rates and completion times.

\subsection{Procedures} \label{proc}
As mentioned earlier, instead of a live experimenter, we used a life-size video/audio 
recording of a experimenter giving instructions. This avatar is the subjects' only source of information 
about the experiment. Actual experimenter involvement 
is limited to strictly off-line activities, 
such as infrequent recalibration of avatar video volume and visual effects, as well as 
occasional repair of some components that suffered minor wear-and-tear damage throughout the study. This unattended setup allows the experiment to run without interruption 24/7 over a 5 month period.

Recall that the central goal of the experiment is to measure performance of subjects who attempt to 
pair their personal Bluetooth device to our Bluetooth device -- an iMAC that uses the SmartBoard 
as an external display. This iMAC is hidden from the subject's view; it is situated directly on the 
other side of the SmartBoard wall in a separate office. During the pairing process, each subject 
is exposed to one randomly selected (from a fixed set) visual stimulus.  This is done by rapid 
change in the ambient lighting of the room's four overhead lightbulbs to the 
chosen stimulus condition.

The experiment runs in four phases: 
\begin{compactenum}
\item Initial: the subject walks in, presses a button on the wall which activates the experiment. 
Duration: instant.
\item Instruction: the avatar delivers instructions via Smartboard display and speakers. 
Duration: 45 seconds.
\item Pairing: the subject attempts to pair personal device with SmartBoard which represents
the hidden iMAC desktop. In this phase, the subject is exposed to one (randomly selected out of 7) 
visual distraction stimulus. Duration: up to 3 minutes.
\item Final: the subject is prompted, on the SmartBoard, to enter some basic demographic 
information, as well as an email address to deliver the reward -- an Amazon discount coupon. 
The information is entered directly into the SmartBoard, acting as a touch-screen input device. 
Duration: up to 6 minutes.
\end{compactenum}
The total duration of the experiment ranged between 5 and 10 minutes.

In order to mitigate any disparities in task completion times between subjects that already had Bluetooth Discovery enabled and those who did not, the avatar informs subjects in the first 15 seconds of the instruction dialog that they will need to perform Bluetooth pairing with their personal device. This gives subjects over 30 seconds to enable Bluetooth Discovery Mode on thier device, if it is not enabled already.

We selected 6 visual effects that differed across two dimensions: color and intensity. 
In terms of color, we picked 3 values in the CIE chromatic space: Red, Blue, and Yellow-Green.
Each is either {\em Solid}, i.e., shown at constant  maximum intensity for the duration of the effect, 
or {\em Flickering}, i.e., its intensity grows and shrinks from the minimum to the maximum and back, 
completing one full cycle every second. In all settings, the maximum saturation was used. 
Color and intensity parameters for the 4 Phillips Hue bulbs under each condition are as follows
(CCV stands for CIE Chromatic Value) \cite{wyszecki1982color}: 
\begin{compactenum}
\item Red, CCV: X= 0.674, Y = 0.322
\item Blue, CCV: X = 0.168, Y = 0.041
\item Yellow-Green, CCV: X = 0.408, Y = 0.517
\item Solid intensity lumen output: 600 lm	
\item Flickering intensity lumen range: 6 lm - 600 lm
\end{compactenum}
These color conditions were picked based on capabilities of programmable bulbs 
as well as background knowledge about emotive effects of color. Phillips Hue
is an LED system based on creating white light. It can not create a blacklight effect
or any achromatic light, which limits color selection to the subspace of the CIE 
color space~\cite{wyszecki1982color} that Hue supports. (See Appendix 2 for more information). 

With that restriction, we looked to the state-of-the-art about emotive reception and sensory 
effects of various colors in the Munsell color space~\cite{1940history}.  
(See Appendix 2 for more information).
It has been shown that \emph{principal hues} -- Red, Yellow, Purple, Blue, and Green
-- are typically positively received. In contrast, \emph{intermediate hues}, i.e., mixtures of 
any two principal hues, are more often negatively associated. Also, various colors have been 
shown to have either an arousing or a relaxing effect on subjects exposed to them. 
Based on this information, we chose three colors that differ as much as possible~\cite{1}: 
\begin{compactitem}
  \item Red: Principal hue with positive emotional connotations, high associated arousal levels
  \item Blue: Principal hue with positive emotional connotations, low associated arousal levels 
  \item Yellow-Green: Intermediate hue with negative emotional connotation,
  high associated arousal levels
\end{compactitem}
Furthermore, we chose to have multiple modalities of light intensity for each color, 
with the expectation that a more complex modality would be more arousing and have a
greater effect than its simple counterpart \cite{koelega1990dynamic}. 
Not having found any previous work on the 
impact of exposure to colored light on performance of security-critical tasks, we include
{\em Solid} light -- the simplest modality of exposure that corresponds  to the base level of 
stimulation. As a more complex modality, we included {\em Flickering}  light. 

Clearly, these two modalities were not the only possible choices. For example, it might have been
intuitive to include even a more complex and startling {\em Strobing} light modality, achievable
through rapid modulation of light intensity. It would have probably engendered a more profound
impact on the subjects. However, ethical considerations coupled with the unattended nature of the
experiment preclude using any modality that could endanger subjects with certain sensitivity 
conditions, such as photosensitive epilepsy. This led us to select a safe flickering frequency of $1 Hz$.

We also found that all three light colors (under both intensity modalities) do not interfere with 
readability of a backlit personal wireless device or the image projected on the Smartboard. 
All experimenters, including one who used corrective lenses, could {\em correctly} read the
screens of their personal devices, under all color conditions and intensity modalities.

\subsection{Prior Results with a Similar Setup}
A very similar setup was used in a previous study that assessed effects 
of unexpected audio distractions on $147$ subjects performing Bluetooth pairing.
As reported in \cite{kaczmarek_unattended_2015}, introduction of audio stimuli 
\emph{significantly increased subject success rates} for every stimulus used. 
There was no significant impact on task completion time for any 
stimulus condition. This phenomenon was likely due
to increased sensory arousal, as discussed in \cite{kaczmarek_unattended_2015}.
Our expectations for the impact of unexpected visual stimuli are rooted in these prior results.

\subsection{Initial Hypotheses} \label{hypo}
We started out by hypothesizing that introduction of unexpected visual distractions during the process
of human-aided pairing of two Bluetooth devices would have similar effects to those observed in prior
experiments with audio distractions. Specifically, we expected two outcomes, as compared to a 
distraction-free setting: \\
\hspace{1.0cm} {\bf [H1]:} Lower error rates, and \\  
\hspace{1.0cm} {\bf [H2]:} No effect on task completion times

\subsection{Recruitment}
The main challenge we encountered in the recruitment process is the scale of the experiments.
Prior studies of usability of human-aided pairing protocols 
\cite{goodrich_using_2009,nithyanand_groupthink:_2010,kainda_usability_2009}, 
demonstrated that 20-25 subjects per tested condition represents acceptable size for obtaining 
statistically significant findings. Our experiment has one condition for each of the six visual 
distraction variations, plus the control condition with no distractions. Therefore, collecting a 
meaningful amount of data requires at least $140$ iterations of the experiment. 

We used a four-pronged strategy to recruit subjects:
\begin{compactenum}
\item Email announcements sent to both graduate and undergraduate 
Computer Science students. 
\item Posters placed (as signboards) near the entrance, and in the lobby, of a 
large campus building which housed the experimental setup. 
\item Several instructors promoted participation in the experiment in
their lectures. 
\item Printed fliers handed out at various campus locations during daily peak pedestrian traffic times.
\end{compactenum}

Recruitment efforts yielded $169$ subjects in total, of whom 125 were male and 44 -- female,
corresponding to a 74\%-26\% gender split.  This is expected, given that
the location of our experimental setup was in the
Computer Science and Engineering part of campus. Most subjects (161) were 
of college age (18-24 years), while 8 were in the 30+ group. This distribution is 
not surprising given the university population and the fact that older subjects 
generally correspond to researchers, faculty and staff, all of whom are much 
less likely to be attracted to being a subject in an experiment. 

As follows from the above, our subjects' demographic was dominated by young, tech-savvy 
male undergraduate students. 

\section{Results}
\label{sec:results}
This section discusses the results, starting with data cleaning and proceeding to
subject task completion effects.

\subsection{Data Cleaning}
\label{subsec:cleaning}
We had to discard subject data for three reasons.

First, although instructions (in fliers, announcements and signs near the setup) specifically 
stated that subjects were to arrive alone, and perform the experiment without anyone else present, 
$37$ groups (2 or more) of subjects participated. We found that the initial participant from each group 
performed in a manner consistent with individual subjects. However, subsequent group members 
who tried the experiment were (not surprisingly) significantly faster and more 
accurate in their task completion. Consequently, we discarded data of every subject who 
arrived in a group and was not the initial participant. We discuss this issue in more detail in Appendix 1.

The second reason for discarding data would have been due to subject auditory and/or visual impairment. 
A subject with an auditory impairment would have difficulties understanding the avatar's 
spoken instructions. A visually impaired subject would have difficulties with using the Smartboard
and with the pairing process which relies on reading and comparing numbers.
After carefully reviewing all subject video records, we could not identify any obvious 
visual or auditory impairment in any subject.

Some subjects successfully completed the experiment several times, perhaps  
hoping to receive multiple participation rewards. This occurred despite explicit 
instructions to the contrary. The system automatically rejected any repeated pairing 
attempts from devices already paired with the system, and any repeated attempts with 
different devices were discovered  by visual inspection of subject trials. Every 
such repeated instance was discarded. 

%

\subsection{Task Failure Rate}
Table \ref{tab:FR} shows the number of subjects who, respectively, 
succeeded and failed at Bluetooth device pairing under each 
stimulus condition. It also details the failure rate for each condition.

\begin{table}[!htb]
\parbox{.45\textwidth}{
\vspace*{-0.3cm}
\smaller
\centering 
\caption{\scriptsize Subject Failure Statistics}
\label{tab:FR}
{\setlength{\extrarowheight}{10pt}
\begin{tabular}{||c|c|c|c||}
\hline \cline{1-4}
Stimulus & \#Successful &\#Failed  & Failure \vspace*{-0.3cm} \\
				&  Subjects & Subjects & Rate
\\ \hline \cline{1-4}
None (control)	& 32 	& 15 	& 0.32  
\\  \hline
Solid Red &	         11 &	9	& 0.45
\\ \hline
Flickering &	9 &	11	& 0.55
\vspace*{-0.3cm}\\
Red & & &
\\ \hline
Solid Blue &	14 &	6	& 0.30
\\ \hline
Flickering &           	8 & 12	 & 0.60
\vspace*{-0.3cm}\\
Blue  & & & 
\\ \hline
Solid &           	10 & 12	 & 0.54
\vspace*{-0.3cm}\\
Yellow-Green & & &
\\ \hline
Flickering &           	7 & 13 & 0.65
\vspace*{-0.3cm}\\
Yellow-Green & & &
\\ \hline
{\bf Total} &           	91 & 78	& 0.46
\\ \hline \cline{1-4}
\end{tabular}}
}
\hfill
\parbox{.45\textwidth}{
\scriptsize
\centering 
\caption{\scriptsize Barnard's Exact Test on failure rates}
\label{tab:BT}
{\setlength{\extrarowheight}{10pt}
\begin{tabular}{||c|c|c|c|c|c||}
\hline \cline{1-6} \bf
Stimulus & Total & Failure & Wald & Nuisance & {\it $p$} 
\vspace*{-0.3cm} \\
			  &  Pairings & Rate &  Statistic & Parameter &  \rm
\\ \hline
None(Control) & 47 & 0.32& --&--&--
\\  \hline
Solid Red &	        20 &	0.45	& 1.02 & 0.88 & 0.17
\\ \hline
Flickering &	20 &	0.55	& 1.77 & 0.86 & 0.04
\vspace*{-0.3cm}\\
Red & & & & &
\\ \hline
Solid Blue &		20 &	0.30	& 0.15 & 0.05 & 0.49
\\ \hline
Flickering &          	20 & 0.60	 & 2.14 & 0.96 & 0.03
\vspace*{-0.3cm}\\
Blue& & & & &
\\ \hline
Solid &           22 & 0.54	 &  1.79 & 0.94 & 0.06
\vspace*{-0.3cm}\\
Yellow-Green& & & & &
\\ \hline
Flickering &           		20 & 0.65  & 2.51 & 0.91 & 0.01
\vspace*{-0.3cm}\\
Yellow-Green & & & & &
\\ \hline \cline{1-6}
\end{tabular}}
}
\end{table}

Table \ref{tab:BT} shows results from Barnard's exact test applied pairwise 
to the subject failure rate of the control condition and each stimulus. It demonstrates 
that differences between failure rates are statistically significant at the $\alpha = 0.05$ 
level with respect to all {\em Flickering} conditions: {\em Flickering Red}, 
{\em Flickering Blue},  and {\em Flickering Yellow-Green}. 
This even holds if we apply a conservative Bonferroni correction to account for three 
pairwise comparisons.  This leads us to the mixed rejection of the initial hypothesis \textbf{H1},
as the failure rate increases significantly with the introduction of certain kinds 
of visual distractions, and remains unaffected by others. The next section 
discusses this further.

Table \ref{tab:Odds} shows odds ratios and 95\% confidence interval for the failure rates under 
each stimulus, as compared to the control condition's failure rate. Interestingly, under this analysis, 
only the confidence intervals of  {\em Flickering Blue} and {\em Flickering Yellow-Green} do not include a 
possible odds ratio of $1.0$. Therefore -- under this method of analysis -- they are the
only statistically significant stimuli at the $\alpha = 0.05$ level. The confidence interval 
defined for the {\em Flickering Red} condition challenges the claim of statistical significance at 
the $\alpha = 0.05$ level, as established by Barnard's exact test.  

We also examined subject failure rates by gender. As shown by Table \ref{tab:gender}
there is no statistically significant difference in failure rates between male and female participants;
Wald statistic $= 0.36$, nuisance parameter = $0.01$, $p = 0.46$.

\begin{table}[!htb]
\parbox{.45\textwidth}{
\smaller
\centering
\caption{\scriptsize Subject Failure Rate by Gender}
\label{tab:Odds}
{\setlength{\extrarowheight}{10pt}
\begin{tabular}{||c|c|c||}
\hline \cline{1-3}
Stimulus &  Odds Ratio  &  95\% Confidence
\vspace*{-0.3cm} \\
	&	wrt control & Interval wrt control
\\ \hline \cline{1-3}
None (control)	& -	& -- 
\\  \hline
Solid Red &	          1.70 &	0.60-5.11
\\ \hline
Flickering Red &	2.61 &	0.89-7.63
\\ \hline
Solid Blue &		0.91 &	0.29-2.85 
\\ \hline
Flickering Blue &           	3.20 & 1.08-9.47 
\\ \hline
Solid Yellow-Green &          1.79 & 0.91-7.24
\\ \hline
Flickering Yellow-Green &           	3.96 & 1.31-11.6
\\ \hline \cline{1-3}
\end{tabular}}
}
\hfill
\parbox{.45\textwidth}{
\small
\centering 
\caption{\scriptsize Subject Failure Rate by Gender}
\label{tab:gender}
{\setlength{\extrarowheight}{10pt}
\begin{tabular}{||c|c|c|c||}
\hline \cline{1-4}
Gender &  \#Successful & \#Unsuccessful  & Failure \vspace*{-0.3cm} \\
			 &  Subjects       & Subjects              & Rate
\\ \hline \cline{1-4}
Male	& 65	& 59	& 0.48	
\\  \hline
Female &	       25&	20	& 0.44
\\ \hline \cline{1-4}
\end{tabular}}
}
\end{table}

\subsection{Task Completion Times}
Table \ref{tab:times}\footnote{Std Dev $=$ Standard Deviation} 
\footnote{DF $=$ Degrees of Freedom.} shows average completion times in successful trials 
under each stimulus. 
After applying a conservative Bonferroni correction to account for six pairwise comparisons between 
individual stimulus conditions and the control condition, every stimulus condition shows an overwhelmingly 
large, statistically significant departure 
from the control condition. This results in rejection of hypothesis \textbf{H2}. The following
section examines possible causes of this slowdown, as well as its implications.

\begin{table}[ht!]
\parbox{.45\textwidth}{
\scriptsize
\centering 
\caption{\scriptsize Avg times (sec) for successful pairing.}
\label{tab:times}
{\setlength{\extrarowheight}{10pt}
\begin{tabular}{||c|c|c|c|c|c||}
\hline \cline{1-6}
{\scriptsize Stimulus} & {\scriptsize Mean}  & {\scriptsize Std} & DF wrt & t-value  &   $p$
\vspace*{-0.3cm} \\
 & Time 	& Dev &  control & wrt control & 
\\ \hline \cline{1-6}
None & 34.50 & 11.93 & -- & -- & -- 
\\ \hline
Solid Red &	        87.81 & 	24.56 & 	41 & 9.56 &  $<0.001$
\\ \hline
Flickering &	90.44 &	15.62 & 39 & 11.59 & $<0.001$
\vspace*{-0.3cm}\\
Red& & & & & 	
\\ \hline
Solid Blue &	106.36 &	17.39 & 44 & 16.32 & $<0.001$ 
\\ \hline
Flickering &           	91.25 & 24.11 & 38 & 9.61 & $<0.001$
\vspace*{-0.3cm}\\
Blue& & & & &
\\ \hline
Solid &           90.30 & 19.08 & 40 & 11.1 & $<0.001$
\vspace*{-0.3cm}\\
Yellow-Green& & & & & 
\\ \hline
Flickering  &           	90.29 & 19.06 & 37 & 10.01 & $<0.001$
\vspace*{-0.3cm}\\
Yellow-Green& & & & &  
\\ \hline \cline{1-6}
\end{tabular}}
}
\hfill
\parbox{.45\textwidth}{
\small
\centering 
\caption{\scriptsize Cohen's $d$ on Completion Times wrt Control}
\label{tab:Cohen}
{\setlength{\extrarowheight}{10pt}
\begin{tabular}{||c|c||}
\hline \cline{1-2}
Stimulus &  Cohen's d  
\vspace*{-0.3cm} \\
	&	wrt control 
\\ \hline \cline{1-2}
None (control)	& -
\\  \hline
Solid Red &	        -3.42
\\ \hline
Flickering Red &	-4.49
\\ \hline
Solid Blue &	-5.33 
\\ \hline
Flickering Blue &  -3.90
\\ \hline
Solid Yellow-Green &  -4.12 
\\ \hline
Flickering Yellow-Green &   -4.29 	
\\ \hline \cline{1-2}
\end{tabular}}
}
\end{table}

Table \ref{tab:Cohen} shows Cohen's $d$ for completion times under each stimulus when compared to 
the control condition. $|d| > 1.0$ in all cases, which means that every stimulus condition shows an 
overwhelmingly large, statistically significant departure from the control condition for the evaluation of 
Cohen's $d$. This result is statistically significant: it indicates that, with convincing probability, the 
mean completion time observed under the control is representative of a different distribution than 
that observed under each stimulus condition. This supports rejection of hypothesis \textbf{H2}.

Next, we looked into subject completion times for successful completion attempts by gender.
Results are displayed in Table \ref{tab:gendertimes}. A pairwise t-test shows that 
observed differences are not statistically significant; $t(84)=0.04$, $p=0.96$. 

\begin{table}[ht!]
\parbox{.45\textwidth}{
\small
\centering 
\caption{\scriptsize Avg times (sec) by gender}
\label{tab:gendertimes}
{\setlength{\extrarowheight}{10pt}
\begin{tabular}{||c|c|c||}
\hline \cline{1-3}
Gender & Mean  & Standard 
\vspace*{-0.3cm} \\
  & Time    & Deviation 
\\ \hline \cline{1-3}
Male & 75.27 & 22.31
\\ \hline
Female & 75.20 & 24.10
\\ \hline \cline{1-3}
\end{tabular}}
}
\hfill
\parbox{.45\textwidth}{
\small
\centering
\caption{\scriptsize One-Way ANOVA test} 
\label{tab:ANOVA2}
{\setlength{\extrarowheight}{10pt}
\begin{tabular}{||c|c|c|c|c|c||}
\hline \cline{1-6}
 & Sum of  & DF &Mean &$F$ & $p$\vspace*{-0.3cm}\\
 & Squares   & & Square& & 
\\ \hline \cline{1-3}
Between & 2964.28 & 5 & 592.86 & 1.466 & 0.217 \vspace*{-0.3cm}\\
Groups& &&&&
\\ \hline
Within & 21440.33 & 53 & 404.535 & &  \vspace*{-0.3cm}\\
Groups&&&&&
\\ \hline
Total & 24404.61 & 58 & & &
\\ \hline \cline{1-6}
\end{tabular}}
}
\end{table}

Finally, we preformed Bartlett's test for homogeneity of variances as well as a One-Way 
analysis of variance (ANOVA) test between average task completion times of all stimulus conditions, excluding the control. 
Bartlett's test failed to reject the null hypothesis that all stimulus conditions share the same 
variance ($\chi^2 = 2.80$, $p = 0.731$). Furthermore, the one-way ANOVA test indicated no 
significant difference between any sample distributions (F = 1.466, p = 0.217.) 
Table \ref{tab:ANOVA2} shows the results; their implications are discussed in the following section. 

\section{Discussion of Observed Effects} 
\label{sec:discussion}
Several types of visual stimuli appear to have a negative effect 
on the subjects' successful completion of the Bluetooth Pairing task. However, 
collected data shows that this is not consistent across all stimuli. Instead, the 
negative effect may be tied to certain features of the particular stimulus. Instances 
of significant degradation in subject success rates were linked to the {\em Flickering} modality, for 
all color stimuli. This result implies that emotional perception of the stimulus may not be as much 
of a contributing factor to the overall increase of subject arousal as the presence of a 
dynamic visual stimulus. Also, in contrast with a previous study of audio distractions 
that observed positive effects \cite{kaczmarek_unattended_2015}, 
we noted no benefit to subject success rates under {\bf any} visual stimulus.

These negative and neutral responses to static and dynamic light stimuli, respectively, 
are reinforced by the psychological concept of attentional selectivity. This concept assumes 
that the capture of an individual's attention by an aversive stimulus is likely to be momentary, 
occurring primarily when the stimulus is first introduced.  In cognitive science, attention is 
conceptualized as a limited resource. For good evolutionary reason, the greatest demand on 
attention is in response to any change in one's environment. Once an assessment of the stimulus 
is made, and determined not to require additional action, attentional devotion to that stimulus 
fades quickly. This means that -- while a static, adverse lighting change may remain adverse
throughout its duration -- its capacity to interfere with subject performance will fade rapidly after 
its onset. Instead, dynamically changing stimuli can more effectively capture subject 
attention and impair their performance, since many assessments are needed for 
many environmental changes occurring throughout the stimulus's duration.

Negative impact on subject task completion rates prompts a new attack vector for the 
adversary who controls ambient lighting. By taking advantage of color effects 
with shifting intensity levels, the adversary could force a user into failing  
Bluetooth pairing as a denial-of-service (DoS) attack. Moreover, 
the adversary might induce failure by using positively perceived colors of varying intensity. 
These colors may not even register as malicious in the user's mind, 
as they are innately associated with beneficial or pleasant emotions.

However, a much greater effect was observed in terms of average completion time. During 
review of subject trials, we noted that, upon exposure to the stimulus, subjects often 
take their gaze off their personal device (or the avatar) and focus their 
attention to the colorful, and possibly flickering, lights. The resulting delay frequently caused the 
subject's device to exit the Bluetooth pairing menu due to a time-out, and re-initiate the 
pairing protocol, resulting in much longer completion times overall. 

Furthermore, as shown by Table \ref{tab:ANOVA2}, the introduced delay in subject task completion time was 
not based on the particular stimulus. Instead, the mere presence of a visual stimulus 
was enough to slow down successful subjects. Similar to the result in inducing user failure, 
the adversary is not forced to rely on an overtly malicious stimulus in order to cause 
substantial slowdown in task completion.  However, the adversary has even more choices
in stimulus selection, since all stimuli (including those with static intensity levels) 
were shown to impact task completion times the same way.

This effect shows further power for the adversary in control of ambient lighting. 
One possibility is that the adversary's goal is a denial-of-service attack by frustrating user's 
pairing attempts. In a more sinister scenario
the adversary could try to ``buy time'' by introducing its own malicious device(s)
alongside changes to ambient lighting and then leverage the user's lapse in focus 
(when being exposed to new sensory stimuli) to trick the user into pairing with that device. 
In the worst case, the adversary might take advantage of the user's inattentiveness while their 
gaze shifts away from their device and trick them into accepting a non-matching 
authenticator.

\section{Unattended Setup: Limitations}
\label{sec:lessons}
Based on our earlier discussion of Data Cleaning,  some subjects' data had to be removed 
from the dataset because they did not conduct the experiment alone.
This occurred even though all recruitment materials (and means)  as well as the avatar's
instructions stated that subjects were to perform the task alone. This illustrates a basic limitation 
of the unattended setup: no one is present to enforce the rules in real time.\footnote{However, 
it would have been possible (though quite difficult in practice) to instrument our recording of the experiment to abort upon 
detecting simultaneous presence of multiple subjects.} 

We did not manage to capture fine-grained data about the subjects' awareness of a distraction. 
We have some anecdotal evidence from video recordings showing that some subjects noticed 
the distraction in obvious ways, e.g., verbal remarks or turning their heads. However, we have no
evidence of subjects who failed to notice the stimulus. Information about subjects noticing
a change in the environment is very important to the development of a realistic adversary model 
for future studies.


\section{Study Shortcomings}
\label{sec:limits}
In this section we discuss some shortcomings of our study.

\subsection{Homogeneous Subjects}
Our subject group was dominated by young, tech-savvy male college students. 
This is a consequence of the experiment's location. Replication of our 
experiment in a non-academic setting would be useful. However, recruiting a really diverse group of 
subjects is hard. Ideal venues might be stadiums, concert halls, fairgrounds or shopping malls. 
Unfortunately, deployment of our unattended setup in such public locations is logistically infeasible. 
Since these public areas already have many sensory stimuli, reliable adjustment of our 
subjects' arousal level in a consistent manner would be very hard. Furthermore, it would be
very difficult to secure specialized and expensive experimental equipment.


In addition to being tech-savvy, young subjects are in general more apt to quickly recover
from changes in the lighting of their surroundings than older adults \cite{kline_vision_1985}.
It is possible that unexpected visual stimuli would have a 
different effect on an older (less technologically adept) population.

\subsection{Sufficiently Diverse Stimuli} \label{divstim}
We selected six conditions to obtain as 
many diverse stimuli types as we could rigorously test, in addition to control. 
We first varied them by changing the regularity of the stimulus, 
expecting that a varying signal would have greater impact on 
subjects' arousal than a steady signal. We then varied the colors, with 
the expectation that using colors that evoked different emotive responses 
and general arousal levels would impact task performance differently. 

An ideal experiment would have included a stimulus with negative emotional 
connotation and low arousal levels. However, between three colors,
two intensity conditions, and the control, we had seven total conditions to test. 
Furthermore, due to the nature of our experiment, we could only reasonably 
expect each subject to be tested under a single condition, since prior knowledge about		 	
the experiment would clearly bias the results. 
Adding just one additional stimulus (for both intensity modalities) would have required at least 
40 more subjects. This would have placed a heavy logistical burden for our 
already nearly-depleted subject pool.

We also note that variance in intensity of our flickering modality did not approach the 
technical limit of Philips Hue bulbs. Instead, we deliberately limited the frequency of 
intensity fluctuations to $1 Hz$ in order to avoid any possible negative reaction from 
light-sensitive subjects. This ethical issue does not reflect real-world conditions 
where an adversary (with no ethical qualms) could create a very fast strobing effect,
possibly causing physical harm.

\subsection{Synthetic Environment}
Our unattended setup, while a step closer to an everyday setting than a sterile and highly controlled lab, 
is still quite synthetic. First, our choice to place it in a low-traffic area makes it quieter than many common 
settings. Second, our choice to situate it indoors makes it free of temperature fluctuations, air flow, and 
exposure to sunlight. Finally, our equipment (such as the Smartboard projector system) is not 
commonly encountered by most subjects.

\subsection{Ideal Setting} \label{ideal}
Drawing upon aforementioned shortcomings, the ideal setting for our experiment would be one where:
\begin{compactitem}
\item Subject demographics are more varied
\item Subjects are not aware of the nature of the experiment until they are debriefed after task completion
\item The environment is more commonplace
\item The task is more security-critical
\end{compactitem}
All of these criteria could be trivially met if, for example, we conducted the experiment at a 
busy bank ATM. The task at hand would be the obviously security-critical entry 
of the subject's PIN. A modern ATM comes standard with all of the features needed for our experiment:
it has a keypad, a screen, a speaker (for visually impaired users), a video camera, and are in 
areas that are artificially lit. Similarly, a busy gas station would fit our needs, as each fuel pump 
typically includes a keypad for PIN entry, speakers, a screen, artificial lighting, and a video camera 
recording the transaction. However, despite their attractive qualities, 
there would be serious ethical and logistical obstacles to
setting up an unattended automated experiment in one these location examples.


\section{Conclusions \& Future Work}
\label{sec:future}
As human participation in security-critical tasks becomes more commonplace, so does the incidence
of users performing these tasks while subject to accidental or malicious distractors. This strongly 
motivates exploring user error rates and their reactions to various external stimuli.
Our efforts described in this paper shed some light on understanding human errors in 
security-critical tasks by studying the effects of visual stimuli on users attempting 
to pair two Bluetooth devices.

We feel that this unattended experiment paradigm is a valuable approach that deserves further study. 
The development of standardized unattended and automated experimental setups could greatly lower the 
logistical and financial burdens associated with conducting large-scale user studies.

Given the observed negative effect on subject completion times, one interesting next step 
would be to conduct a similar experiment, where, instead of measuring
subjects' ability to pair Bluetooth devices, we would examine the rates of incorrect pairing when
the subjects are shown mis-matched numbers during the pairing process. 
This could help us determine whether (and how) visual distractions make users more likely to 
pair their device to some other (perhaps adversary-controlled) device.

Another direction is investigating effects of hybrid (e.g., audio/visual) distractions. 
Finally, we plan to conduct a study of subjects performing security-critical tasks,  while being
exposed to {\em multiple visual stimuli} lasting longer than 3 minutes. This might allow us
to learn whether subjects' sensory arousal is the result of the surprise (due to the sudden 
visual stimulus), or an unavoidable psychophysical reaction.

\section*{Acknowledgments}
\label{sec:ack}
This research has been supported by NSF Grant CNS-1544373. 
\bibliographystyle{abbrv} 
\bibliography{vision-sec}

\appendix
\section*{Appendix 1: Analysis of Group Initiators}
\label{app:Init}

\begin{table}[h]
\parbox{.45\textwidth}{
\small
\centering 
\caption{{\scriptsize Failure Rates: Initiators vs. Individuals}}
\label{tab:group}
{\setlength{\extrarowheight}{10pt}
\begin{tabular}{||c|c|c|c||}
\hline \cline{1-4}
Participant Type &  \#Successful & \#Unsuccessful  & Failure \vspace*{-0.3cm} \\
			 &  Subjects       & Subjects              & Rate
\\ \hline \cline{1-4}
Group Initiator	& 19	& 18	& 0.49
\\  \hline
Individual &	       72&	60	& 0.45
\\ \hline \cline{1-4}
\end{tabular}}
}
\hfill
\parbox{.45\textwidth}{
\small
\centering 
\caption{\scriptsize Avg times (sec): Initiators vs. Individuals}
\label{tab:grouptimes}

{\setlength{\extrarowheight}{10pt}
\begin{tabular}{||c|c|c||}
\hline \cline{1-3}
Participant  & Mean  & Standard \vspace*{-0.3cm}\\
  Type & Time    & Deviation 
\\ \hline \cline{1-3}
Group Initiator & 76.63 & 23.00
\\ \hline
Individual & 76.20 & 17.93
\\ \hline \cline{1-3}
\end{tabular}}
}
\end{table}	

We considered potential differences in failure rates between subjects 
who performed the task alone, and those who did it as part of a group. 
As mentioned in the discussion of Data Cleaning, for each group, 
we only consider the initial participating group member, referred to as the Group Initiator. 
As Table \ref{tab:group} shows, there is no significant difference between failure rates of 
individual subjects and Group Initiators; Wald Statistic $= 0.34$, 
Nuisance parameter $= 0.01$, $p = 0.51$. Furthermore, as Table \ref{tab:grouptimes} shows, 
a pairwise t-test of completion times for individuals -- compared to group initiators -- 
shows that observed differences are not statistically significant; $t(84)=0.09$, $p = 0.93$.

\section*{Appendix 2: A Few Colorful Words}
\label{app:colors}
\subsection*{Munsell Color System}
\begin{figure}
    \centering
    \begin{minipage}{0.45\textwidth}
        \centering
\includegraphics[height=2.0in, width =\textwidth]{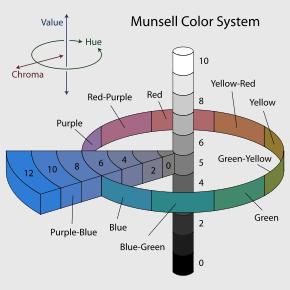}
\caption{{\small {\small  Munsell Color Space (Image best viewed in color}}}
\label{fig:Munsell}
\end{minipage}
\end{figure}

The Munsell Color System is used for creating and describing colors. In it, 
all colors are grouped into two categories: primary and intermediate hues. 
Primary hues include: Red, Yellow, Purple, Blue, and Green, arranged in a circular 
shape as in Figure \ref{fig:Munsell}. Intermediate hues are mixtures of two adjacent 
primary hues, such as Yellow-Green or Purple-Blue. Colors are defined on three 
dimensions: hue, lightness, and color purity. The Munsell system is based on human 
perception which makes it useful for rigorously defining human reaction to specific color forms. 
However basing the system on human perception makes the Munsell system a poor tool for 
direct conversion of light described by its physical wavelength into human-perceptible color.

\subsection*{CIE Color Space}

\begin{figure}	
    \centering
    \begin{minipage}{0.45\textwidth}
        \centering
\includegraphics[height=2.0in, width =\textwidth]{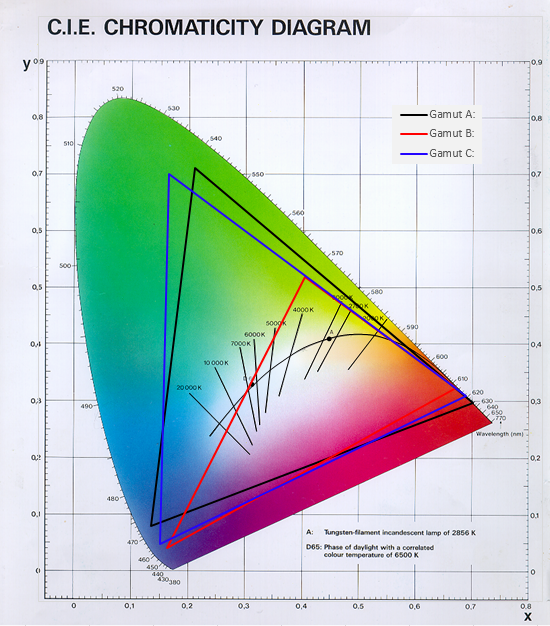}
\caption{{\small {\small  Phillips Hue CIE Color Space (Image best viewed in color}}}
\label{fig:chrom}
\end{minipage}
\end{figure}

The Phillips Hue bulbs use the CIE color space. In CIE, 
colors are defined as a 2-dimensional space with X and Y values moving along a roughly 
triangular curve that corresponds to the translation of wavelengths of light to their human 
perception in the visible spectrum. The exact color range of the Philips Hue bulb is 
shown in Figure \ref{fig:chrom}

\section*{Appendix 3: Unattended Experiment Setup} \label{app:pics}

\begin{figure}
    \centering
    \begin{minipage}{0.45\textwidth}
        \centering
\includegraphics[height=2.0in, width =\textwidth]{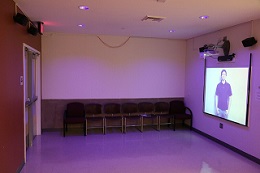}
\caption{{\small The Experiment Environment during the Solid Blue condition (Image best viewed in color)} }
\label{fig:proxy}
    \end{minipage}\hfill
    \begin{minipage}{0.45\textwidth}
        \centering
\includegraphics[height=2.0in, width =\textwidth]{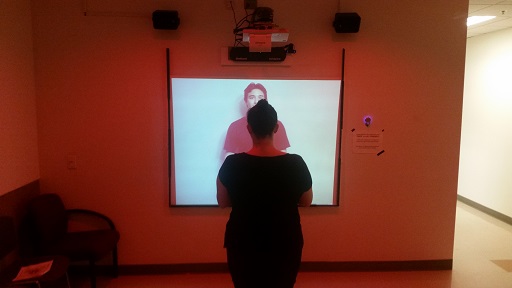}
\caption{{\small The subject's perspective during the Solid Red condition (Image best viewed in color.)}}
    \end{minipage}
\end{figure}

\begin{figure}
    \centering
    \begin{minipage}{0.45\textwidth}
        \centering
\includegraphics[height=2.0in, width =\textwidth]{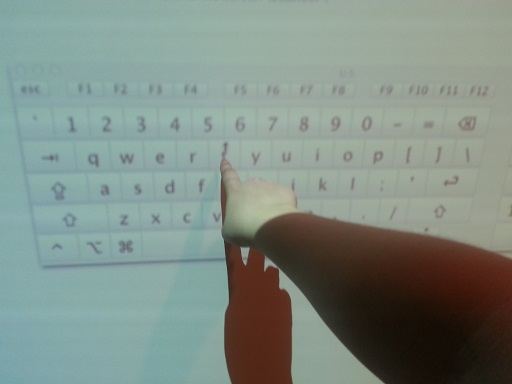}
\caption{{\small  Subject entering email address on Smartboard} }
\label{fig:proxy}
    \end{minipage}\hfill
    \begin{minipage}{0.45\textwidth}
        \centering
\includegraphics[height=2.0in, width =\textwidth]{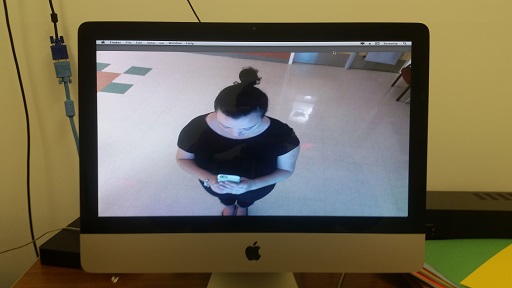}
\caption{{\small Post-experimental review of video recordings (separate office)}}
    \end{minipage}
\end{figure}


\end{document}